\documentclass[conference]{IEEEtran}

\usepackage{hyperref}
\hypersetup{
     colorlinks = true,
     citecolor = green,
     }

\ifCLASSINFOpdf
  \usepackage[pdftex]{graphicx}
\else
\fi

\usepackage{framed}
\usepackage{subcaption}

\usepackage{xcolor}
\definecolor{cs-blue}{rgb}{0.035, 0.114, 0.235}
\definecolor{turq}{rgb}{0.365, 0.835, 0.851}

\usepackage{tcolorbox}

\tcbset{
  takeawaybox/.style={
    colback=turq!7!white, 
    colframe=turq, 
    coltitle=black, 
    fonttitle=\bfseries\small, 
    title=Key Takeaways: 
  }
}

\usepackage{balance}
\balance

\begin{document}

%
\title{Industrial Code Quality Benchmarks: Toward Gamification of Software Maintainability}


\author{
\IEEEauthorblockN{Markus Borg}
\IEEEauthorblockA{\textit{CodeScene and Lund University}\\
Malmö, Sweden \\
markus.borg@codescene.com}
\and
\IEEEauthorblockN{Amogha Udayakumar}
\IEEEauthorblockA{\textit{CodeScene}\\
Malmö, Sweden \\
amogha.udayakumar@codescene.com}
\and
\IEEEauthorblockN{Adam Tornhill}
\IEEEauthorblockA{\textit{CodeScene}\\
Malmö, Sweden \\
adam.tornhill@codescene.com}
}

%


\maketitle

\begin{abstract}
Software maintainability is essential for long-term success in the software industry. Despite widespread evidence of the high costs associated with poor maintainability, market pressure drives many organizations to prioritize short-term releases. This focus leads to accumulating technical debt worldwide. In this preliminary work, we propose maintainability gamification through anonymous leaderboards to encourage organizations to maintain a sustained focus on code quality. Our approach envisions benchmarking to foster motivation and urgency across companies by highlighting thresholds for leaders and laggards. To initiate this concept, we analyze a sample of over 1,000 proprietary projects using CodeHealth scores. By examining the distribution of these scores across various dimensions, we assess the feasibility of creating effective leaderboards. Findings from this study offer valuable insights for future design activities in maintainability gamification.
\end{abstract}


%

\section{Introduction} \label{sec:intro}
Inadequately managed Technical Debt (TD) is a silent killer in the software industry. Numerous studies report how its deteriorating effects can lead to long-term problems in software projects~\cite{lenarduzzi_systematic_2021,ernst_technical_2021}. A systematic literature review by Li \textit{et al.} identified that primary studies have reported that TD has compromised all seven top-level qualities in ISO~25010~\cite{li_systematic_2015}, e.g., maintainability, security, performance efficiency, and reliability. By far, most primary studies highlight the detrimental effects on maintainability. In this light, we use the following pragmatic TD definition to guide the work in this paper: ``TD is code that is more expensive to maintain than it should be''~\cite{tornhill_your_2024}.

Maintainability does not receive the attention it deserves in the software industry. Martini \textit{et al.} found that only 7\% of organizations systematically manage TD~\cite{martini_technical_2018}, e.g., identification, prioritization, repayment, and continuous monitoring. When it comes to repayment, the most common approach mentioned in the literature is by far \textit{refactoring}, followed by rewriting, automation, and reengineering~\cite{li_systematic_2015}. Our empirical studies have illuminated the costs associated with TD~\cite{tornhill_code_2022,borg_u_2023} and, more recently, demonstrated the value of highly maintainable code~\cite{borg_increasing_2024}. While organizations recognize the value, it often fades under the pressure of urgent business needs, which leads to prioritizing new feature development over TD remediation. We posit that novel approaches to motivate a sustained focus on TD are needed.

As comparisons can be motivational, benchmarking might serve as a vehicle for positive change. For example, a systematic literature review in the healthcare sector shows how benchmarking encouraged quality improvement~\cite{willmington_contribution_2022}. If properly designed using validated KPIs, we believe that the same effects could apply in the software industry. On the organizational level, benchmarking might motivate a strive for reaching the KPIs associated with industry \textit{leaders}. Vice versa, finding yourself in the lower range in industry, i.e., among the \textit{laggards}, might create a sense of urgency leading to increased refactoring budgets from the management level. Furthermore, we hypothesize that there might be a virtuous cycle on the individual level, i.e., a positive feedback loop, as the developer happiness and TD presence are connected~\cite{graziotin_unhappiness_2017} and happy developers are more productive~\cite{graziotin_are_2013}.

We propose \textit{maintainability gamification} by creating anonymous leaderboards across companies. Leaderboards are one of the most commonly discussed game elements in gamification of Software Engineering (SE)~\cite{barreto_gamification_2021}, and we believe it would be suitable for our motivational purposes. To increase the relevance of the benchmarking results, we assume software organizations will want to filter the results in different ways, e.g., based on industry segment, project size, or programming language. In this preliminary work, we pave the way for cross-company maintainability leaderboards guided by three research questions:

\begin{itemize}
    \item[RQ\textsubscript{1}] What is the distribution of maintainability in proprietary projects in the software industry?
    \item[RQ\textsubscript{2}] Can we observe systematic differences across clusters of industry segments?
    \item[RQ\textsubscript{3}] How can other ways of data segmentation be used for leaderboard creation? 
\end{itemize}

The remaining sections of the paper cover a short background to gamification and benchmarking, the research method, results and discussion, and conclusions with ideas for future work.

\section{Background} \label{sec:bg}
Many studies discuss various approaches to SE gamification. Padreira \textit{et al.} cover the early work between 2011 and 2014 in a systematic review~\cite{pedreira_gamification_2015}. Based on 29 primary studies, the authors found that ``system implementation'' is the ISO/IEC/IEEE 12207~\cite{international_organization_for_standardization_systems_2017} process area that has been considered the most. In a more recent secondary study, Barreto and Franca identified 130 primary studies~\cite{barreto_gamification_2021}. They report that ``software construction'' was the software area that had been researched the most. Moreover, the most commonly used game elements were 1) points, 2) badges, and 3) leaderboards. 

In a more comprehensive review, Porto \textit{et al.} included 103 primary studies~\cite{porto_initiatives_2021}. They again found that development (equivalent to implementation and construction) was the primarily addressed target. While the authors argue that the field is immature, some primary studies indicate that gamification can promote goals such as adherence to best programming practices~\cite{barik_perspective_2016,foucault_fostering_2019} and regular refactoring~\cite{arai_gamified_2014,elezi_game_2016}. 

Several gamification projects have targeted educational settings~\cite{alhammad_gamification_2018}. Dubois and Tamburrelli reported their experience using SE gamification in Italian university courses~\cite{dubois_understanding_2013}. In their setup, some student groups could compare their static code analysis results from SonarQube to those of others, thus stimulating competition. Their findings suggest that access to competitive benchmarking led to a slightly better average among these groups. In the same spirit, other researchers have proposed using gamification to incentivize TD management in industry practice~\cite{foucault_gamification_2018,crespo_role_2022}.

We build on previous work on SE gamification. Like many before us, we focus on software development/implementation/construction -- especially the code quality dimension \textit{maintainability}. Furthermore, we use points and leaderboards as commonly proposed in previous work~\cite{barreto_gamification_2021}. Moreover, we use Garcia \textit{et al.}'s GOAL framework~\cite{pedreira_gamification_2015} to structure our ideas in Sec.~\ref{sec:fw}.

Industry benchmarking to identify leaders and laggards is a common practice across different topics. Yasin provides a comprehensive review of the history of benchmarking in industry~\cite{yasin_theory_2002}. The review describes how benchmarking originates in the two predecessors: competitive analysis and quality function deployment. The review highlights the seminal contributions of Camp, widely regarded as the father of benchmarking, whose work at Xerox Corporation drove significant improvements in production efficiency and cost reduction~\cite{camp_benchmarking_1989}.

Benchmarking is also used in SE. Already in 2003, Sim \textit{et al.} called for increased focus on SE benchmarking in a well-cited ICSE paper~\cite{sim_using_2003}, largely motivated by its success in computer science. More recently, Hasselbring presented an empirical standard for benchmarking~\cite{hasselbring_benchmarking_2021} that is now included in the ACM SIGSOFT Empirical Standards initiative. Between these papers, there are examples of benchmarking studies that target SE performance metrics. For example, Cragg conducted benchmarking of ICT practice, competence and performance in small companies~\cite{cragg_benchmarking_2002}. He identified six practices that differentiate leaders from laggards. 
Similarly, Pai \textit{et al.} benchmarked 79 projects at a CMMI level~5 organization, employing data envelopment analysis -- a promising approach for future work, as discussed in Sec.~\ref{sec:fw}
~\cite{pai_benchmarking_2015}.

When presenting quantitative benchmarking results, it is common to highlight leaders and laggards. We believe this is particularly valuable to drive action in our gamification application. Where to actually draw the lines is up to us as designers, and it should be based on the effects one hopes to achieve. In this paper, we regard the top 10\% of projects as leaders, and the bottom 10\% as laggards. In the figures that follow in Sec.~\ref{sec:res}, we indicate the 10th and 90th percentiles.

\section{Method} \label{sec:method}

\subsection{Data Collection and Preprocessing}
We selected a subset of proprietary projects from customers with at least 10 employees. The total dataset contains roughly 1,000 projects, which we consider feasible for this preliminary study. For confidentiality reasons, we cannot disclose further details of the sampling process.

We ran complete CodeScene analyses for all projects. A project contains a codebase that can be hosted in one or several git repositories. In this work, we focus on analyzing source code from a maintainability perspective. Our primary metric CodeHealth has recently outperformed SonarQube, the dominant static code analysis tool in industry, on the Maintainability Dataset~\cite{borg_ghost_2024} established by Schnappinger \textit{et al.}~\cite{schnappinger_defining_2020}.

In this paper, we report two CodeHealth scores. First, the average CodeHealth of a codebase. The average is weighted by the number of Source Lines of Code (SLoC) per file, i.e., lines of code excluding comments and blank lines. Second, we present the Hotspot CodeHealth, the weighted average CodeHealth for the files that have changed the most frequently over the past year. We say that code smells in these files carry the highest TD interest.

We complemented the projects with metadata from the customer relationship management platform HubSpot\footnote{www.hubspot.com}, i.e., number of employees and classification into industry segments. However, the latter data is available for less than 20\% of the customers. Similarly, far from all analyses in this sample of projects contain all dimensions used for data segmentation. This is due to both technical and licensing-related constraints, which will not be elaborated upon. As a result, the total number of projects represented in the figures in Sec.~\ref{sec:res} varies. We accept this limitation and report as much information as possible -- meaningful patterns can still be identified.

\subsection{Data Analysis and Visualization}
We rely on HubSpot's categorization of companies into industry segments. In August 2024, HubSpot used 147 different categories, out of which 32 are represented in our dataset. We subsequently used OpenAI's GPT-4o model to organize the 32 industry segments into three clusters:

\begin{enumerate}
    \item[C-A] Consumer and Hospitality Industries, incl. consumer goods, gambling casinos, hospitality, real estate, retail, apparel fashion, and leisure travel tourism.
    \item[C-B] Professional Services and Education, incl. human resources, consumer services, education management, staffing and recruiting, civil engineering, marketing and advertising, insurance, investment management, financial services, professional training coaching, legal services, primary secondary education, public safety, and design.
    \item[C-C] Industrial and Technological Segments, incl. computer software, transportation trucking railroad, oil energy, information technology and services, automotive, mechanical or industrial engineering, telecommunications, construction, printing, biotechnology, and medical devices.
\end{enumerate}

To complement the cluster analysis, we provide several other approaches to data segmentation. Thresholds were selected to create three groups of roughly similar size, as follows:

\begin{enumerate}
    \item \textbf{Codebase size}
    \begin{itemize}
        \item Small: SLoC $\leq$ 20,000
        \item Medium: 20,000 $<$ SLoC $\leq$ 80,000
        \item Large: SLoC $>$ 80,000
    \end{itemize}
    \item \textbf{Company size}
    \begin{itemize}
        \item Small: 10 $\leq$ employees $\leq$ 100
        \item Medium: 100 $<$ employees $\leq$ 350
        \item Large: $>$ 350 employees 
    \end{itemize}
    \item \textbf{Codebase age}
    \begin{itemize}
        \item Greenfield: repo created $\geq$ 2022
        \item Brownfield: 2018 $\leq$ repo created $\leq$ 2021
        \item Legacy: repo created $<$ 2018
    \end{itemize}
    \item \textbf{Programming language.} We report file-level results for the top 8 languages in the dataset.
\end{enumerate}

For each dimension under study, we analyze both raw frequencies and frequencies weighted by the SLoC codebase size. For the raw numbers, each project counts equally much regardless of its characteristics. For the weighted numbers, larger projects contribute more, with their impact scaled according to their relative size.

All visualizations in Sec.~\ref{sec:res} rely on Probability Density Functions (PDF). This design decision brings two main advantages. First, PDFs preserve confidentiality by omitting absolute frequencies. Second, PDFs enhance visual comparisons by creating continuous distribution shapes, allowing for easier identification of patterns through kernel smoothing.

\section{Results and Discussion} \label{sec:res}
For each figure that follows in this section, we use the same semantics. All curves show PDFs of the data, illustrating the relative likelihood of different values occurring within each distribution. Yellow curves represent the raw distribution of CodeHealth across projects. Blue curves represent the distribution of CodeHealth weighted by projects' SLoC. We highlight the mode for each PDF, i.e., the point where the PDF reaches its maximum value, indicating the most probable value in the distribution. Furthermore, for both curves, we highlight the 10th and 90th percentiles to indicate the thresholds for ``laggards'' and ``leaders,'' respectively.

\subsection{RQ\textsubscript{1}: Maintainability Distribution in Industry} \label{sec:rq1}
\begin{figure}[t]
    \centering
    \begin{subfigure}{\columnwidth}
        \centering
        \includegraphics[width=\textwidth]{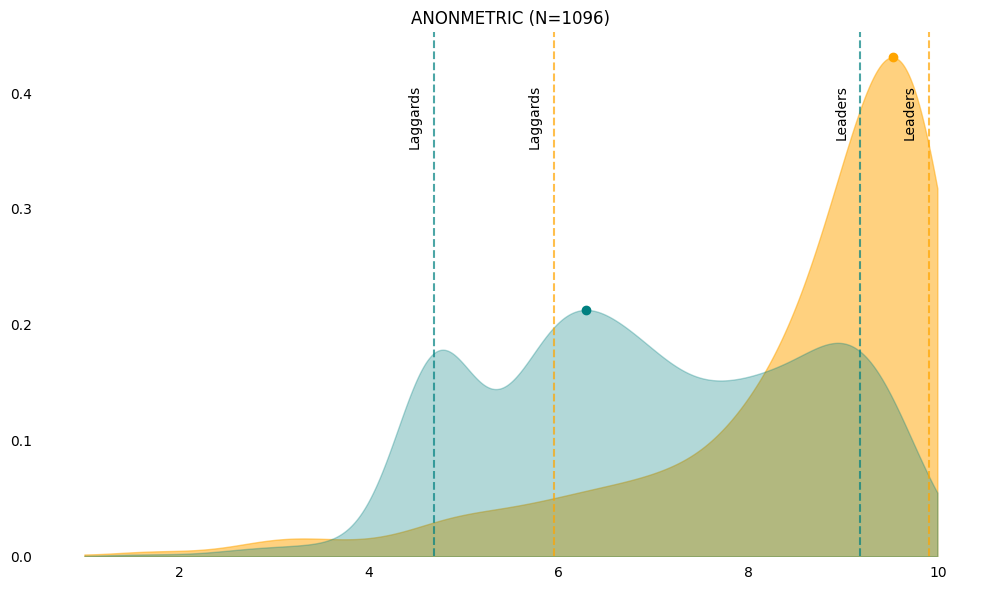} 
        \caption{Average CodeHealth in the dataset.}
        \label{fig:avg_tot}
    \end{subfigure}
    \vspace{5mm}
    \begin{subfigure}{\columnwidth}
        \centering
        \includegraphics[width=\textwidth]{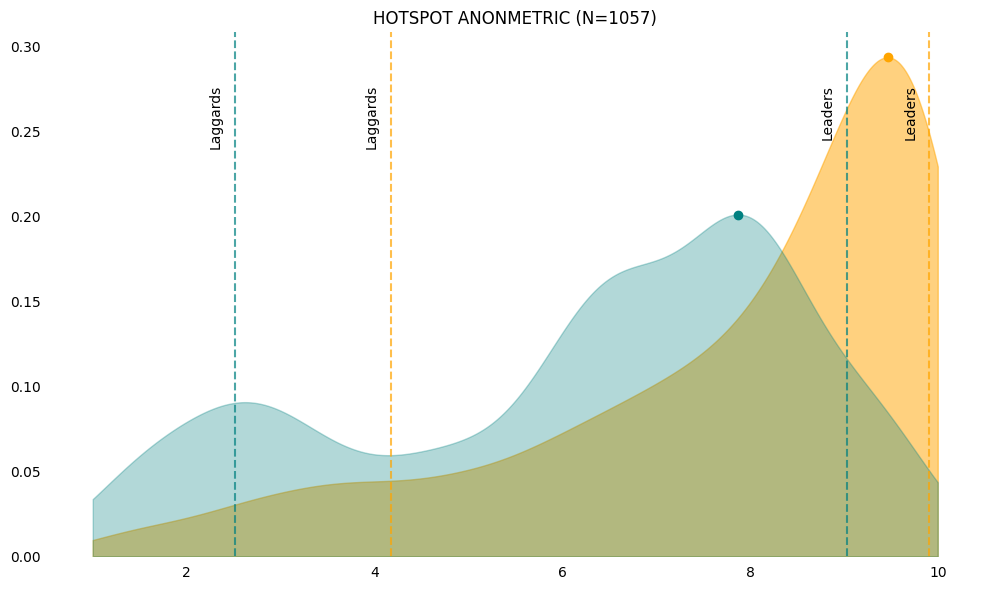} 
        \caption{Hotspot CodeHealth in the dataset.}
        \label{fig:avg_hs_tot}
    \end{subfigure}
    \caption{Distribution of CodeHealth in the dataset.}
    \label{fig:tot}
\end{figure}

Fig.~\ref{fig:avg_tot} shows the distribution of average CodeHealth for all 1,000+ projects in our dataset. We observe that a vast majority of projects are in the interval above 8 with a mode at 9.5. However, there is a long tail of projects with lower average CodeHealth and the laggards threshold can be seen at 5.9. The size-weighted blue curve in the figure shows a contrasting distribution, for which the probabilities are rather uniform between the leaders and laggards. 

Fig.~\ref{fig:avg_hs_tot} presents the distribution of Hotspot CodeHealth for all applicable projects, i.e., the maintainability of the most actively changed files in the projects. The raw distribution is similar to the average CodeHealth, but the long tail is heavier -- it contains more probability mass. This means it is more common for projects to have a low Hotspot CodeHealth compared to a low average CodeHealth. The laggard threshold is found at 4.1, which is close to the CodeHealth alert level (4)~\cite{tornhill_code_2022}. The blue curves shift the thresholds further left, which shows that large projects have less maintainable Hotspots. This finding is in line with Lehman's first and second laws of software evolution, that is 1) systems tend to grow more complex over time and 2) systems' maintainability tends to decrease unless efforts are actively taken to mitigate it. Maintaining a large system with a Hotspot CodeHealth around 2 must be a challenging endeavor, leaving little time for new feature development.

The results suggest that the raw distributions of maintainability scores for proprietary projects are heavily skewed with a majority of the projects in the upper quality interval. Due to the dominant probability mass around high scores, the leaderboard positions around the leaders might appear volatile. That is, small CodeHealth changes might cause big moves. From a gamification perspective, future work must evaluate whether this is motivating or if it rather feels arbitrary for participating organizations. When the PDF is weighted by SLoC, the distributions become wider -- which could potentially be advantageous for leaderboard creation.

\begin{tcolorbox}[takeawaybox, title=RQ\textsubscript{1}: Distribution of maintainability in industry?]
Maintainability distributions tend toward higher values but are left-skewed due to many projects with low CodeHealth scores. Larger projects generally show lower maintainability.
\end{tcolorbox}

\subsection{RQ\textsubscript{2}: Benchmarking per Industry Cluster} \label{sec:rq2}
\begin{figure*}[t]
    \centering
    \begin{subfigure}{\columnwidth}
        \centering
        \includegraphics[width=\textwidth]{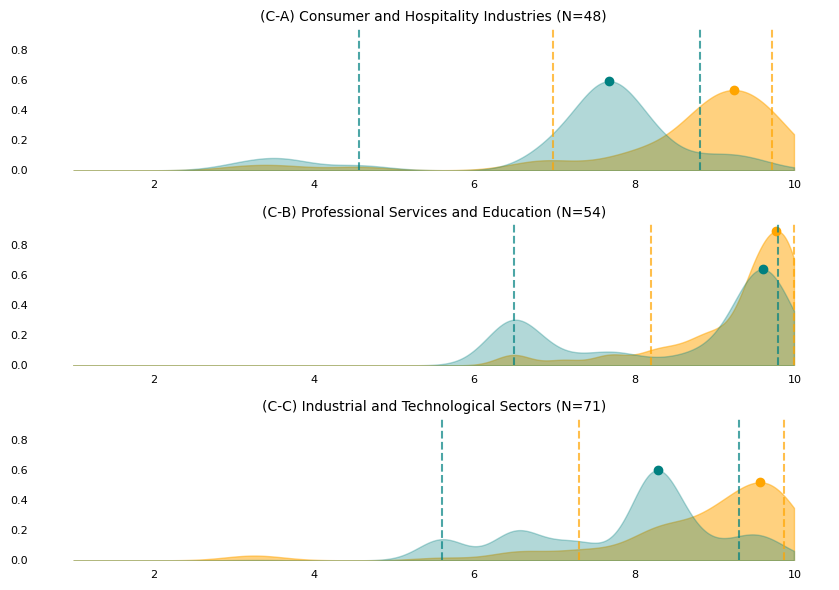} 
        \caption{Average CodeHealth per industry sector cluster.}
        \label{fig:avg_ch_cluster}
    \end{subfigure}
    \begin{subfigure}{\columnwidth}
        \centering
        \includegraphics[width=\textwidth]{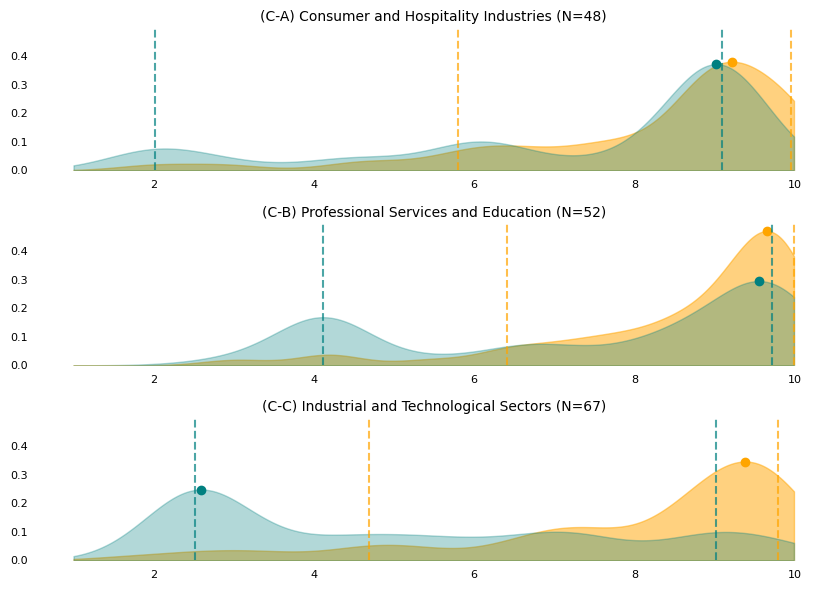} 
        \caption{Hotspot CodeHealth per industry sector cluster.}
        \label{fig:avg_hs_ch_cluster}
    \end{subfigure}
    \caption{Maintainability benchmarking per industry sector cluster.}
    \label{fig:clusters}
\end{figure*}

Fig.~\ref{fig:clusters} shows results from benchmarking CodeHealth for three clusters of industry sectors. Comparing the raw average CodeHealth scores (yellow distributions) in Fig.~\ref{fig:avg_ch_cluster}, we see similar patterns across the clusters. Most of the studied projects appear to be well-maintained with modes for C-A, C-B, and C-C at 9.2, 9.8, and 9.6, respectively. For all clusters, there are tails of projects with lower average CodeHealth scores.

Looking at the CodeHealth distributions with project contributions weighted by size, we see a shift towards lower scores -- as was the case for RQ1. This is particularly evident for clusters C-A and C-C, for which the density peaks appear at 7.3 and 8.3, respectively. For C-B, the distribution turns bimodal, with a lower peak appearing at 6.6, whereas the upper peak remains largely unchanged. The results show that larger projects contain more files that are hard to maintain, which we further discuss in relation to RQ3.

Fig.~\ref{fig:avg_hs_ch_cluster} shows the distribution of Hotspot CodeHealth for the three clusters. Again, we find that the general patterns for the yellow distributions are similarly left-skewed across the three clusters. Mode values around are high, between 9.2 and 9.6. When investigating the size-weighted distributions, we see considerable shifts for clusters C-B and C-C. For C-B, the distribution turns bimodal, with a lower peak appearing around CodeHealth 4.1. For cluster C-C, the shift is more extreme --- the distribution is almost inverted, with a peak at 2.5 -- a very low score -- and a heavy uniform tail towards the higher ranges. The finding is alarming, as this low maintainability level for the files that change the most is a strong indication of high-interest TD. Maintaining and evolving such projects will be severely hampered by substandard code quality. 

We hypothesize that cluster C-C (Industrial and Technological Segments) contains more legacy systems, which might shift results to the lower range. While C-A (Consumer and Hospitality Industries) displays a common long-tail distribution for Hotspot CodeHealth, projects in C-B (Professional Services and Education) appear to have a large subset of hard-to-maintain projects. We analyze this further in RQ3 as we explore other data segmentations.

\begin{tcolorbox}[takeawaybox, title=RQ\textsubscript{2}: Differences between industry segment clusters?]
The raw maintainability distributions across the three clusters look similar. However, when weighed by project size, notable differences emerge, especially for Hotspot CodeHealth.
\end{tcolorbox}

\subsection{RQ\textsubscript{3}: CodeHealth for Other Splits} \label{sec:rq3}
In this section, we compare the maintainability of proprietary projects using other data segmentations.   

\subsubsection{Segmentation by Project Size}
\begin{figure*}[t]
    \centering
    \begin{subfigure}{\columnwidth}
        \centering
        \includegraphics[width=\textwidth]{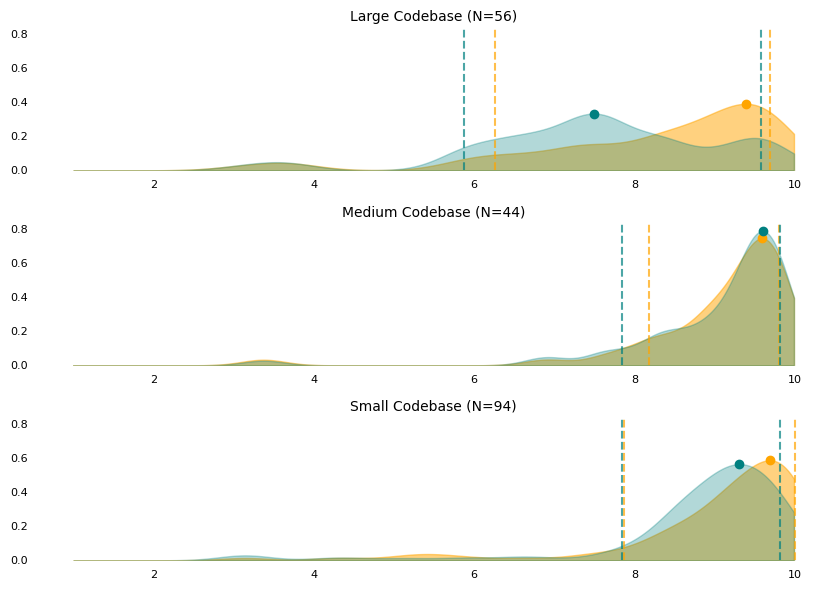} 
        \caption{Average CodeHealth grouped by project size.}
        \label{fig:avg_ch_size}
    \end{subfigure}
    \begin{subfigure}{\columnwidth}
        \centering
        \includegraphics[width=\textwidth]{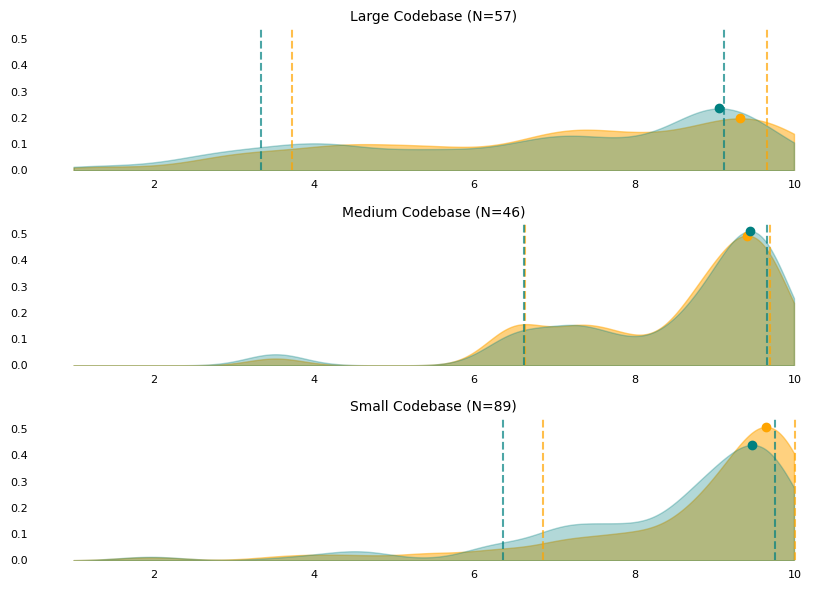} 
        \caption{Hotspot CodeHealth grouped by project size.}
        \label{fig:avg_hs_ch_size}
    \end{subfigure}
    \caption{Maintainability grouped by project size.}
    \label{fig:sizes}
\end{figure*}

Fig.~\ref{fig:sizes} presents CodeHealth scores grouped by project sizes. As expected, weighting the contributions by project size, i.e., the blue curves, has less impact when the data has already been segmented by size. 

For average CodeHealth, we see how the scores decrease with increasing project sizes. This is in accordance with the observations we presented for size-weighted distributions for RQ1 and RQ2. For Hotspot CodeHealth, we see that the spread of the distribution increases for large projects. We observe almost a uniform distribution between 3 and 10, which suggests that some organizations have found sustainable ways to maintain large software projects. Studying lessons learned from the leaders of this group would be a promising source for extracting good maintainability practices. 

\subsubsection{Segmentation by Company Size}
\begin{figure*}[t]
    \centering
    \begin{subfigure}{\columnwidth}
        \centering
        \includegraphics[width=\textwidth]{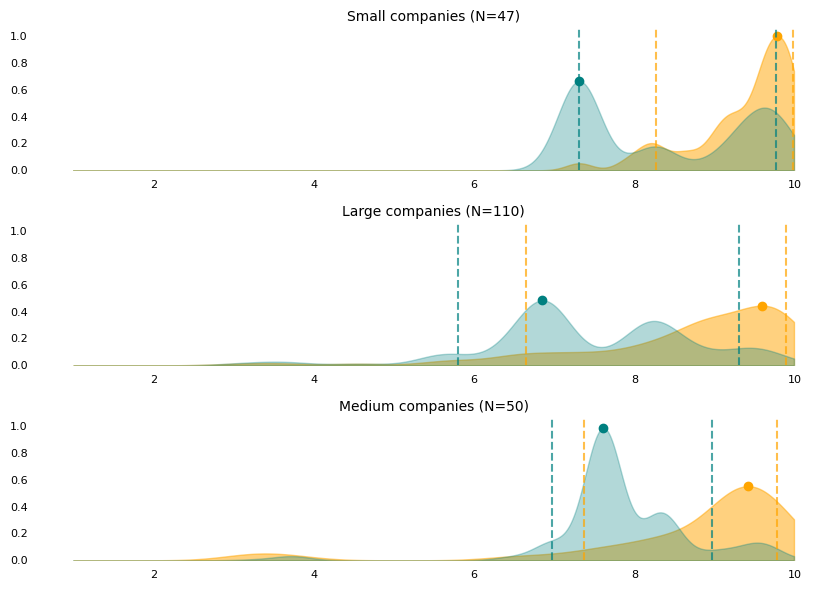} 
        \caption{Average CodeHealth grouped by company size.}
        \label{fig:avg_ch_company}
    \end{subfigure}
    \vspace{5mm}
    \begin{subfigure}{\columnwidth}
        \centering
        \includegraphics[width=\textwidth]{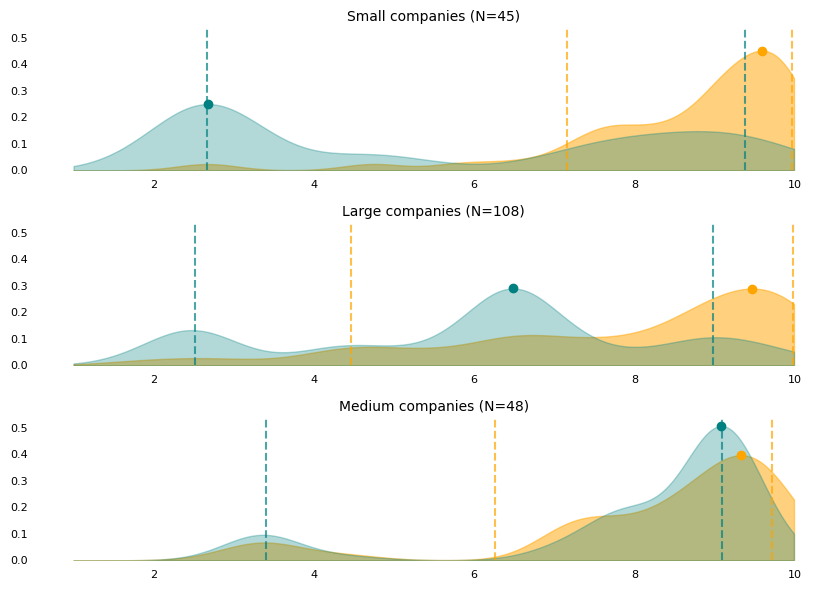} 
        \caption{Hotspot CodeHealth grouped by company size.}
        \label{fig:avg_hs_ch_company}
    \end{subfigure}
    \caption{Maintainability grouped by company size.}
    \label{fig:organizations}
\end{figure*}

Fig.~\ref{fig:organizations} depicts CodeHealth scores grouped by the size of the companies. For the raw numbers in yellow in Fig.~\ref{fig:avg_ch_company}, we see roughly equally shaped distributions, although the spread increases with company size. This shows that the diversity in terms of maintainability increases with organizational size, in line with the increased diversity we observed for larger projects.

Looking at the distribution for size-weighted projects, we see substantial shifts for all segments. For large companies, we again see an increased spread with two peaks at 6.9 and 8.2, respectively. For small companies, the mode is at 7.3, but there is also a peak at 9.6 -- which again suggests diversity. For medium companies, the shift leads to a mode of 7.6 and a right-skewed distribution. 

Fig.~\ref{fig:avg_hs_ch_company} shows distributions of Hotspot CodeHealth. As for average CodeHealth, the distributions look largely the same with a slightly increasing spread for increasing company size. When weighting by the project size, we see a shift toward lower scores for small and large companies. The shift is the most pronounced for small companies, with a peak at 2.7. This is a very low score, which strongly suggests that the data contains large projects from small companies that have accumulated challenging levels of TD in the files that change the most. For large companies, we observe another bimodal distribution, i.e., the mode is at a problematic CodeHealth of 6.5, but there is also a peak at the alarming level of 2.5.

\subsection{Segmentation by Codebase Age}
\begin{figure*}[t]
    \centering
    \begin{subfigure}{\columnwidth}
        \centering
        \includegraphics[width=\textwidth]{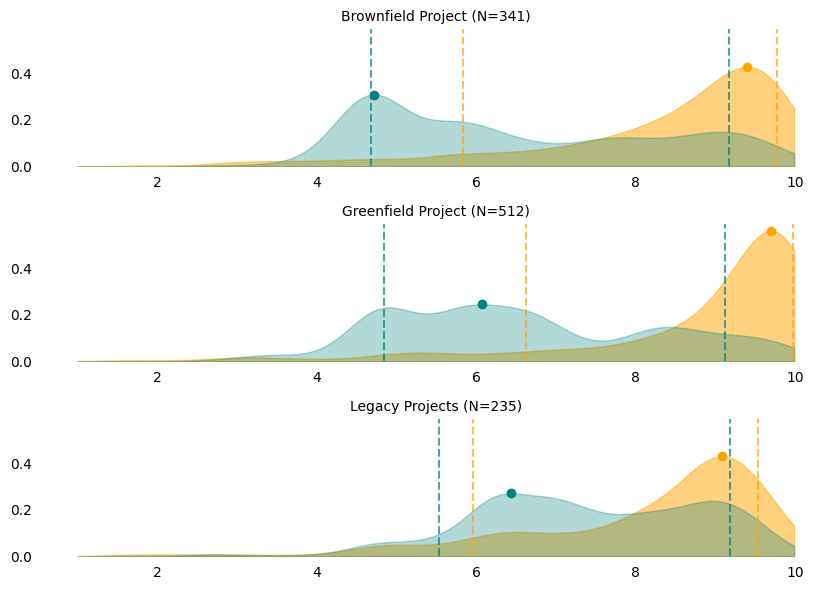} 
        \caption{Average CodeHealth grouped by codebase age.}
        \label{fig:avg_ch_age}
    \end{subfigure}
    \begin{subfigure}{\columnwidth}
        \centering
        \includegraphics[width=\textwidth]{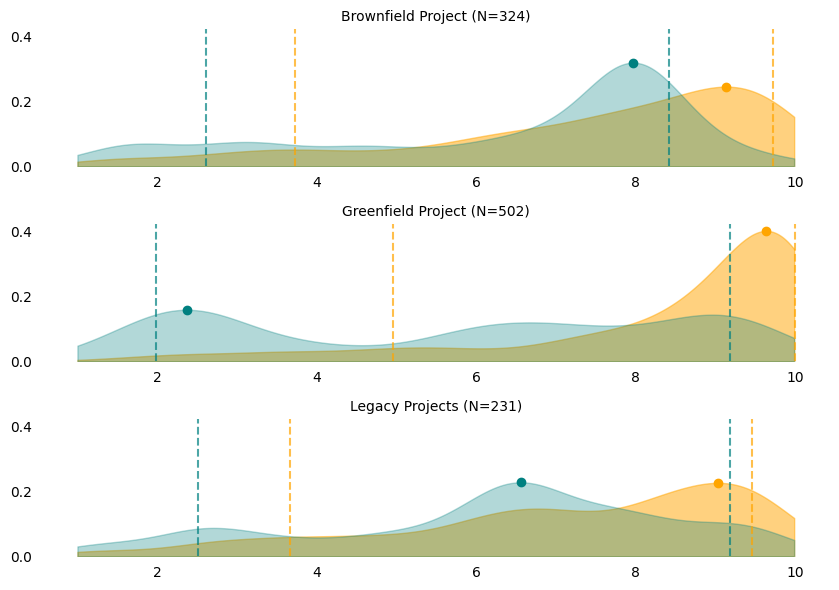} 
        \caption{Hotspot CodeHealth grouped by codebase age.}
        \label{fig:avg_hs_ch_age}
    \end{subfigure}
    \caption{Maintainability grouped by codebase age.}
    \label{fig:age}
\end{figure*}

Fig.~\ref{fig:age} depicts CodeHealth scores grouped by the age of codebases. For raw values, Fig.~\ref{fig:avg_ch_age} shows increasingly pronounced left tails for legacy and brownfield projects compared to greenfield projects. Moreover, the modes of the distributions shift slightly leftward with age. This is an expected finding, as older codebases have more time to diversify during evolution. However, when weghted by project size, all distributions shift considerably to the left. This shows that there are large projects that are likely hard to maintain in all age segments -- including the recently initiated ones.

Fig.~\ref{fig:avg_hs_ch_age} shows distributions of Hotspot CodeHealth. Looking at the raw data in yellow, we find more spread for brownfield and legacy projects -- again an expected finding. When weighting by project size, the distributions shift left as we have discussed for previous plots. The biggest change is visible for greenfield projects, where the mode is as low as 2.4 and the distribution is rather uniform across the entire maintainability interval.

\subsection{Segmentation per Programming Language}
\begin{figure}
    \includegraphics[width=0.5\textwidth]{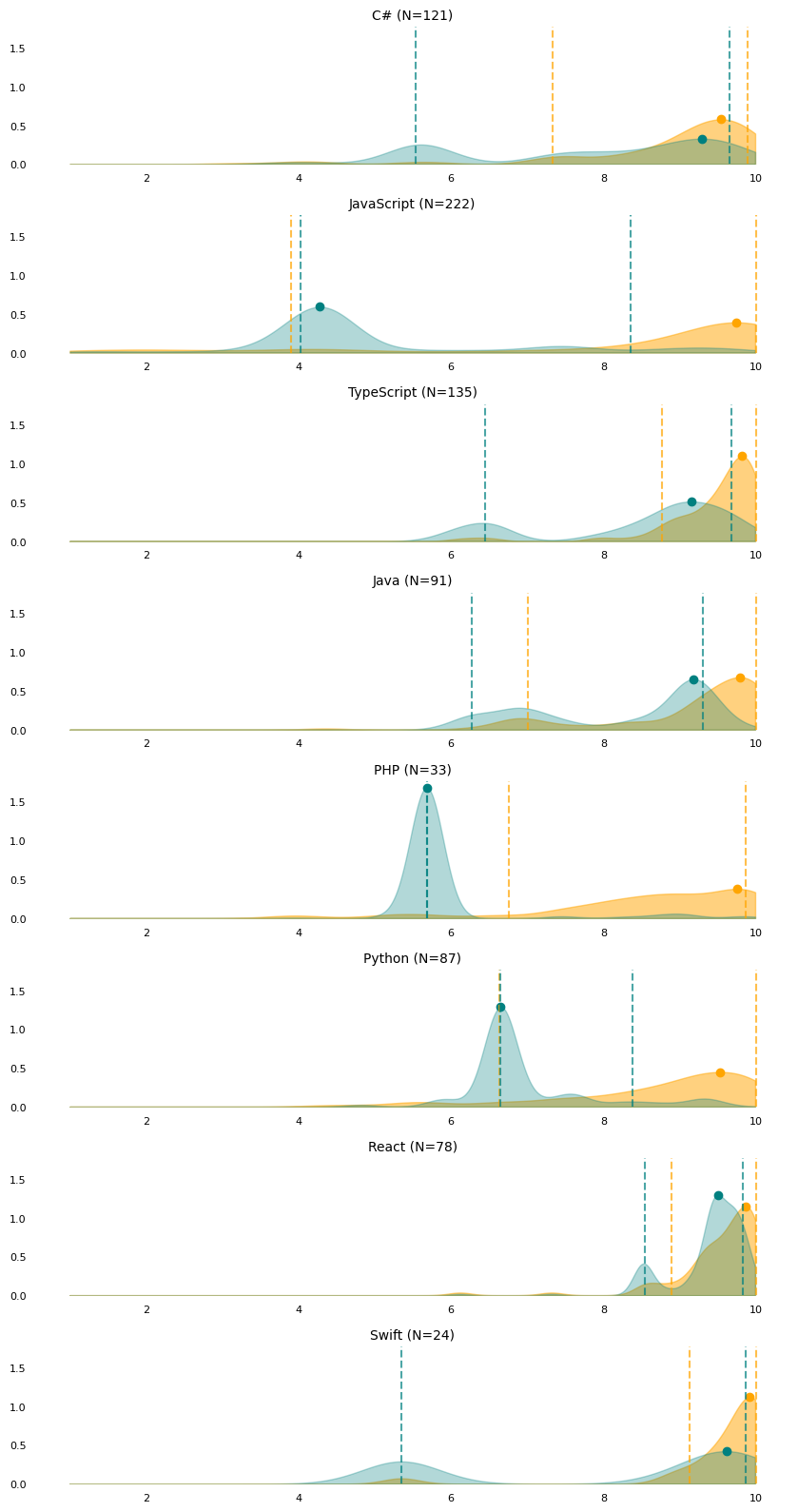} 
    \caption{Maintainability grouped by programming languages.}
    \label{fig:language}
\end{figure}

Fig.~\ref{fig:language} shows average CodeHealth scores for files aggregated by programming language at the project level. Each data point represents a single project's weighted average score for all files of that language. We find that all raw distributions have modes between 9.5 and 9.9 -- suggesting that it is possible to develop maintainable systems in all eight languages covered. While we refrain from drawing conclusions related to languages in this initial study, we do observe some interesting patterns for the size-weighted distributions. First, all distributions turn more or less bimodal when weighing for size. This suggests that there is a general trend in industry that large and problematic files exist no matter the language. Furthermore, it is well-known that there is a strong negative correlation between SLoC and maintainability on the file level~\cite{borg_ghost_2024} -- our new results corroborate this. Second, three languages stand out by having greatly left-shifted modes, i.e., JavaScript (4.3), PHP (5.7), and Python (6.7). Thus, our results suggest that very large and unmaintainable files are more common in these three languages.

\begin{tcolorbox}[takeawaybox, title=RQ\textsubscript{3}: Use other ways of data segmentation?]
Different data segmentation methods can yield meaningful leaderboard rankings, each valuable to different stakeholders. To support diverse perspectives, we recommend enabling interactive data filtering.
\end{tcolorbox}

\section{Limitations and Threats to Validity} \label{sec:threats}
The main objective of this preliminary study is to pave the way for maintainability gamification using leaderboards. As a step in this direction, we study the maintainability characteristics of a large sample of proprietary projects. We realize that a side effect of our initial work is an \textit{ad hoc} survey of industry practice -- few academic studies cover as many proprietary studies as ours. Still, we acknowledge that our sample is not representative of all proprietary software development projects. While there is diversity in terms of domains and languages represented, we believe the sample from anonymous customers is biased toward organizations that pay particular attention to maintainability -- this might shift our distributions toward higher CodeHealth values. Moreover, we know that embedded software is underrepresented in the sample. Had our goal been survey research, we would have prioritized increasing external validity.

Maintainability is an established SE construct, defined for example in ISO~25010. Regarding construct validity, we must consider to what extent we correctly and completely measure it. We rely on the CodeHealth score, which has been validated for correctness in previous studies. Regarding completeness, we acknowledge that it only considers code, while maintainability is also influenced by factors such as documentation and organization. Furthermore, CodeHealth is based on a set of code smells. There are other smells that might influence maintainability, but we argue that we have covered the most impactful ones.

Finally, we discuss some threats to the conclusions we draw regarding the feasibility of future gamification. Much of the paper discusses the characteristics of distributions, and we hypothesize that more uniform distributions are preferable for leaderboard creation. First, the distributions we report depend on the sample of projects. Second, our chosen data segmentations are certainly decisive. We used HubSpot categories for industry segments and ChatGPT for organizing them. The thresholds selected were rather arbitrarily chosen to balance groups. Future work could involve a sensitivity analysis for the threshold values.

\section{Implications for Maintainability Gamification} \label{sec:fw}
Based on our initial analysis of the problem, we structure our future gamification endeavor using the high-level steps of the GOAL (Gamification focused On Application Lifecycle management) framework~\cite{garcia_framework_2017} The framework makes several key design decisions explicit, which both helps us communicate our ideas and organize the work ahead of us. At this point, we focus on the first four steps of GOAL.

\begin{enumerate}
    \item \textbf{Identify the objectives of the gamification.} Encourage a sustained focus on maintainability management, for which success will be measured longitudinally using Hotspot CodeHealth.
    \item \textbf{Player analysis.} The players are companies that compete against others with similar characteristics, e.g., industry segment or codebase size. Whereas it is not the current focus, later hierarchical gamification would be possible where CodeHealth is collected for divisions, departments, projects, or teams -- down to the file level.
    \item \textbf{Scope Definition and Feasibility Study.} The SE scope of the game is from the start of the software construction for as long as the software is being maintained. Within an organization, the game is primarily \textit{collaborative} in line with the ``shared code'' principle of agile software development. Between companies, the game is \textit{competitive}. This preliminary paper is part of a bigger feasibility study.
    \item \textbf{Game Analysis and Design.} The main \textit{game component} is a leaderboard with anonymous entries populated by organizations ranked by points based on CodeHealth. The \textit{game mechanic} is to remove code smells in an organization's source code, ideally in the parts of code that change the most. The \textit{game economy} follows the calculations of CodeHealth scores. No additional \textit{game rules} will be enforced, as the organization's development policies should be sufficient. An example of cheating would be to remove large parts of code to get rid of code smells, but that is not a realistic misuse case. The \textit{aesthetics} will be an interactive online leaderboard with several options for filtering the data -- as indicated by Figs.~\ref{fig:clusters}-\ref{fig:language}.
    \item Development of the Gamified SE Platform. The development will follow our established processes and is not further elaborated here.
    \item Managing, Monitoring, Measuring. Once deployed, the gamified platform will be periodically monitored to analyze the leaderboard and the distribution of industry benchmarking scores.
\end{enumerate}

\section{Conclusions and Future Work} \label{sec:conc}
We hypothesize that gamification through industry benchmarking can effectively motivate organizations to sustain a maintainability focus. Our results show that CodeHealth scores tend toward higher values but exhibit a strong left skew. This distribution suggests that using CodeHealth for leaderboards could create a potentially volatile experience for high-performing ``leaders'' compared to lower-scoring ``laggards.'' While we found no major differences between industry sector clusters, there are many other ways to segment the data to foster engaging comparisons among similar organizations.

Once we have established a maintainability benchmarking dataset, data envelopment analysis would be a promising direction for future work. This type of analysis is used in operations research when evaluating units with several inputs and outputs, where simplistic efficiency metrics fall short. The approach constructs an efficiency frontier, i.e., a boundary of optimal performance, represented by the most efficient units in the benchmarking dataset. Next, each unit gets an efficiency score between 0 and 1 representing how far they are from the frontier. The approach has been used in SE research before, e.g., for comparing the performance of open-source software projects~\cite{wray_evaluating_2008}. As data envelopment analysis is a comparative method, it should fit our ambitions of gamifying maintainability in industry well.

\section*{Acknowledgment}
This work was partly funded by the NextG2Com Competence Centre -- Next-Generation Communication and Computing Infrastructures and Applications -- under the Vinnova grant number 2023-00541.


\bibliographystyle{IEEEtran}
\bibliography{gamify}

\end{document}